\documentstyle[twoside,fleqn,espcrc2,epsfig]{article}

\newcommand{\AmS}{{\protect\the\textfont2
  A\kern-.1667em\lower.5ex\hbox{M}\kern-.125emS}}
\def\kreis{\raise0.85pt\hbox{$\scriptscriptstyle\bigcirc$}}
\def\vollk{\lower0.95pt\hbox{\Large $\bullet$}}
\def\bbox{\lower0.75pt\hbox{$\Box$}}
\def\plus{\raise0.45pt\hbox{$+$}}
\def\diamo{\lower0.45pt\hbox{$\Diamond$}}

\title{
\vspace{-2.0cm}                         % for preprint
{\normalsize DESY 98--160}     \\[-0.2cm]      % for preprint
{\normalsize HLRZ 1998--60}   \\[-0.2cm]      % for preprint
{\normalsize October 1998}      \\              % for preprint
\vspace{0.7cm} 
Topical Results on Lattice Chiral Fermions in the CFA\thanks{Talk 
given by V. Bornyakov at Lattice 1998.}}

\author{V. Bornyakov\address{Institute for High Energy Physics, 
        RU-142284 Protvino, Russia}, 
        A. Hoferichter\address{Deutsches Elektronen-Synchrotron DESY and NIC, 
        D-15735 Zeuthen, Germany},
        G. Schierholz$^{\rm b,}$\address{Deutsches Elektronen-Synchrotron DESY,
        D-22603 Hamburg, Germany}, and
        A. Thimm$^{\rm b,}$\address{Institut f\"ur Theoretische Physik, Freie
        Universit\"at Berlin, D-14195 Berlin, Germany}}

\begin{document}

\begin{abstract}
We report new results on the lattice regularization of the chiral Schwinger
model and the chiral U(1) model in four dimensions in the CFA.  
\end{abstract}

\maketitle

\section{INTRODUCTION}
 
The continuum fermion approach (CFA) to regularizing chiral fermions appears
to be a promising method. For a summary of results obtained so far
see~\cite{kyoto}. The basic idea of the approach 
is the following. We start from a lattice with spacing $a$. We call
this the original lattice. On this lattice
the simulations are done. Next we construct a finer lattice with lattice
spacing $a_f$, using a suitable interpolation of the gauge field~\cite{GKSW}. 
On this lattice we formulate the fermions. The action for a single fermion of
chirality $\epsilon_\alpha$ is taken to be  
\begin{eqnarray}
S_{\epsilon_\alpha} &\!\!\!=\!\!\!& \bar{\psi}{\cal D}_{\epsilon_\alpha}
\psi \nonumber \\
&\!\!\!\equiv\!\!\!& \frac{1}{2} \bar{\psi}\{\gamma_\mu 
(D_\mu^{\epsilon_\alpha +} + D_\mu^{\epsilon_\alpha -}) \label{LA} \\
&\!\!\!-\!\!\!& \frac{r}{2} a_f  D_\mu^{\epsilon_\alpha +} 
D_\mu^{\epsilon_\alpha -}\} \psi, \nonumber 
\end{eqnarray}
where, restricting ourselves to gauge group U(1), 
\begin{eqnarray}
(D_\mu^{\epsilon_\alpha \pm} \psi)(n) &\!\!\!=\!\!\!& 
\pm \;\frac{1}{a_f}\{[P_{-\epsilon_\alpha} + 
P_{\epsilon_\alpha}(U^f_{\pm \mu}(n))^{e_\alpha}] \nonumber \\
&\!\!\! \!\!\!&\times \;\psi(n \pm \hat{\mu}) - \psi(n)\},
\end{eqnarray}
with $U^f_{\pm \mu}$ being the link variable on the fine lattice,
$e_\alpha$ the
fractional charge, and
$P_{\epsilon_\alpha} = (1+\epsilon_\alpha \gamma_5)/2$.
The effective action is then computed from (\ref{LA}) in the limit 
$a_f \rightarrow 0$, where $a$ is kept fixed. Hence the name continuum
fermions.  

A particular feature of this action -- or better of the
Wilson term we have chosen -- is that the Wilson-Dirac operator ${\cal
D}_{\epsilon_\alpha}$ fulfills
the Ginsparg-Wilson relation~\cite{GW,Betal}:  
\begin{equation}
\gamma_5 {\cal D}_{\epsilon_\alpha} +  {\cal D}_{\epsilon_\alpha}\gamma_5
= r\,a_f {\cal D}_{\epsilon_\alpha}\gamma_5 {\cal D}_{\epsilon_\alpha} 
+ O(a_f^2).
\label{GW}
\end{equation}
This is not a great surprise. The motivation behind our construction
was to find an action which obeys the index theorem -- what, in fact, it
does as we shall see below. The Ginsparg-Wilson 
relation (\ref{GW}) does, however, not guarantee that the theory is
invariant under chiral gauge transformations.

Indeed, the resulting effective action, $W_{\epsilon_\alpha}$, is not
gauge invariant, even in the limit $a_f \rightarrow 0$. But there
exists a local, purely bosonic counterterm $C$, so that
\begin{equation}
\widehat{W}_{\epsilon_\alpha} = \lim_{a_f \rightarrow 0} 
W_{\epsilon_\alpha}^\Sigma, \; W_{\epsilon_\alpha}^\Sigma =
W_{\epsilon_\alpha} + C
\label{eff}
\end{equation}
is invariant under chiral gauge transformations. For the imaginary
part this is only true in the anomaly-free model. The counterterm can
be -- and has been -- computed in perturbation theory.

A further important feature of the action (\ref{LA}) is that it has
a shift symmetry which makes sure that the ungauged fermion with
chirality $-\epsilon_\alpha$ decouples.

In the chiral Schwinger model we found for the topologically trivial
sector of the theory
\begin{equation}
\widehat{W}_{\epsilon_\alpha} = \frac{1}{2} (W_V + W_0) + \mbox{i}\,
\mbox{Im} W_{\epsilon_\alpha},
\end{equation}
where $W_V$ and $W_0$ are the effective actions of the corresponding
vector model and the free theory, respectively. The imaginary part 
of the effective action turned out to be given by the harmonic
part of the gauge field, i.e the toron field, alone, and it could be
computed analytically 
from the gauge field we started with. Thus we arrived at an
action which can be simulated relatively easily on the original lattice.
We believe to find a similar result in more realistic models in four
dimensions. 

After so much of introduction, let us now come to our new results.

\section{CHIRAL SCHWINGER MODEL}

Our first results concern the chiral Schwinger
model. Here we want to test whether the index theorem is
fulfilled. This is a non-trivial task because any topologically
non-trivial gauge field involves at least one singular plaquette. 
Some authors have argued that the CFA would fail this test.
This would be true had we employed the standard gauged or ungauged
Wilson terms. Furthermore, we will investigate
whether our approach reproduces the correct anomaly.

In two dimensions the index theorem says that in a background gauge
field configuration of topological charge $Q$ we should find exactly
\begin{equation}
n_{\epsilon_\alpha} = |Q|\, \theta(\epsilon_\alpha Q)
\end{equation}
zero modes of chirality $\epsilon_\alpha$. For a configuration of,
e.g., charge $Q = +1$ this would mean $n_+ = 1$ and $n_- = 0$. 

To leading order in $a_f$ the Wilson-Dirac operator can be written 
\begin{equation}
{\cal D}_{\epsilon_\alpha} = \not{\!\!D}^{\epsilon_\alpha} 
- a_f \frac{r}{2} \not{\!\!D}^{-\epsilon_\alpha} \!
\not{\!\!D}^{\epsilon_\alpha} + O(a_f^2),
\label{D}
\end{equation}
where $\not{\!\!D}^{\epsilon_\alpha}$ is the average of forward and 
backward derivatives. Being a finite matix, the Dirac operator    
$\not{\!\!\!D}^{\epsilon_\alpha} \equiv 
- \not{\!\!\!D}^{-\epsilon_\alpha \, \dagger}$
has the same number of zero modes as the corresponding vector operator, namely
$n_{\epsilon_\alpha} + n_{-\epsilon_\alpha} = |Q|$, thus violating the
index theorem. But the situation is different for the Wilson term  
$\not{\!\!\!D}^{-\epsilon_\alpha} \not{\!\!\!D}^{\epsilon_\alpha} 
\,\equiv\, (\not{\!\!\!D}^{-\epsilon_\alpha}
\not{\!\!\!D}^{\epsilon_\alpha})^\dagger$. It has 
$n_{\epsilon_\alpha}$ zero modes of chirality $\epsilon_\alpha$ and
none of chirality $-\epsilon_\alpha$, exactly as required by the 
index theorem. For small, but finite $a_f$ the zero 
modes approach the value
\begin{equation}
\left(\frac{a_f}{a}\right)^2 \frac{\pi |Q|}{L^2},
\label{zero}
\end{equation}  
where $L$ is the size of the original lattice.

We consider two configurations of charge $Q$. We denote the
link variables on the original lattice by $U_\mu(s) =
\exp(\mbox{i}\theta_\mu(s)), \, -\pi < \theta_\mu(s) \leq \pi$, where $s_\mu$ 
are the lattice points on the original lattice. The first configuration
is~\cite{SV} ($\mbox{mod}~2\pi$)
\begin{equation}
\begin{array}{lcl}
\theta_1(s) &\!\!\! =\!\!\! & F\,s_2 - \bar{\theta}_1, \\
\theta_2(s) &\!\!\! =\!\!\! & \left\{\begin{array}{lr}
                       -\bar{\theta}_2,& \hspace{-0.4cm} s_2=1,\dots,L-1, \\ 
                        FL\,s_1-\bar{\theta}_2,& s_2=L,
                       \end{array} \right.
\label{SV}
\end{array}
\end{equation}
where $F= 2\pi Q/L^2,~\bar{\theta}_1=\pi(L-1)/L^2$ and
$\bar{\theta}_2=\pi/L^2$.
The second configuration is 
\begin{equation}
\theta_\mu(s)= 2\pi Q \epsilon_{\mu\nu} \partial^-_\nu G(s-\bar{s}),\,
\bar{s} = (L/2,L),
\label{VM}
\end{equation}
where $G$ is the inverse lattice Laplacian. Both configurations
have constant field strength $F$, zero toron field, and for $|Q| = 1$
they have one singular plaquette at $s = (L/2,L)$. The two
configurations are related by 
a periodic (topologically trivial) gauge transformation. In the following 
we shall take $L=6$, and we shall use the interpolation given in~\cite{GKSW}.
 
Both configurations give the same value for the effective action
(\ref{eff}). As the anomaly-free model we took 
$\epsilon_\alpha e_\alpha = -1,-1,-1,-1,2$. This 
indicates that we have gauge invariance
in the background of singular gauge fields as well. For $Q=1$ we find
exactly one zero mode with chirality $+$, and none with chirality
$-$. For $Q=2$ we find two zero modes with chirality $+$, and none
with chirality $-$. For negative charges we find the same result but with
$+$ and $-$ interchanged, in agreement with the index theorem.
The values of the zero modes are in good agreement with the analytical result
(\ref{zero}).

\begin{figure}
\vspace{-0.15cm}
\begin{centering}
\epsfig{figure=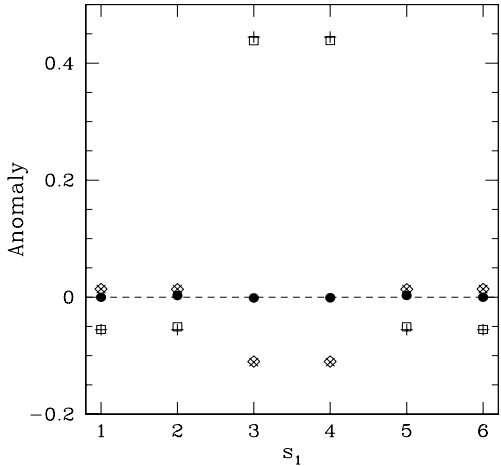,height=6.5cm,width=6.92cm}
\vspace{-0.9cm}
\caption{The anomaly as a function of $s_1$ for $s_2 = 4$. The l.h.s. 
(r.h.s.) of (\ref{anomaly}) is marked by
$\bbox$ ($\plus$) for $\epsilon_\alpha e_\alpha = -1$, and by 
$\Diamond$ ($\times$) for $\epsilon_\alpha e_\alpha = 2$. 
The result for the anomaly-free model is denoted by $\vollk$.}
\vspace{-0.7cm}
\end{centering}
\end{figure}

A further requirement of the method is that
it reproduces the correct anomaly. In the chiral Schwinger model 
the anomaly condition reads
\begin{equation}
\frac{\delta W_{\epsilon_\alpha}}{\delta h(s)} = \mbox{i} \epsilon_\alpha
e_\alpha^2 \frac{1}{4\pi} \epsilon_{\mu\nu} F_{\mu\nu}(s),
\label{anomaly}
\end{equation}
where $h(s)$ is a gauge transformation. We consider configuration (\ref{VM}) 
with $Q=-1$ and $\bar{s} = (L/2,L/2)$ now. The r.h.s. of (\ref{anomaly}) is
known analytically. The l.h.s. is computed for $a/a_f = 5$, i.e. on the 
$30^2$ lattice. In Fig.~1 we show the result for both sides separately for 
$\epsilon_\alpha e_\alpha = -1$ and $2$, as well as for the anomaly-free 
model. We find excellent agreement between our numerical results and the 
theoretical expectations (i.e. the l.h.s. versus the r.h.s.). Moreover, 
we see that the anomaly cancels in the anomaly-free model. We find similar 
results for configuration (\ref{SV}).

\section{CHIRAL U(1) MODEL IN 4-D}

Our other new results are on the chiral U(1) model in four dimensions. 
This model is simpler than the chiral Schwinger model because it
has no topological charge and zero modes to worry about. 

The first question is again whether the theory exists at all, i.e. can be 
made finite by the addition of local counterterms.
We find three possible counterterms: 
\begin{eqnarray}
C_2 &\!\!\!=\!\!\!& c_2 \sum \left(\sum_{\mu} A_\mu^2\right),  \nonumber \\
C_4 &\!\!\!=\!\!\!& c_4 \sum \left(\sum_{\mu} A_\mu^2\right)^2, \\
C_\partial &\!\!\!=\!\!\!& c_\partial \sum \left(\sum_{\mu} 
\partial_\mu A_\mu\right)^2. \nonumber
\end{eqnarray}
As before, the coefficients, $c_2, c_4$ and $c_\partial$, can be computed 
in lattice perturbation theory. The result so far is
\begin{equation}
c_2 = 0.03717\, a_f^{-2}, \;
c_4 = 0.00052.
\label{ct}
\end{equation}
We cannot quote any number for $c_\partial$ yet. Numerically we find that 
gauge invariance is, within our present accuracy, already restored by the 
counterterms $C_2$ and $C_4$, if we use the perturbative values (\ref{ct})
for the coefficients. This indicates that $c_\partial$ is very small, or 
zero. As an upper bound we can quote $c_\partial < 10^{-4}$. 

What distinguishes the chiral theory from the corresponding vector 
theory is the imaginary part, $\mbox{Im}\,W_{\epsilon_\alpha}$, 
of the effective action. In two dimensions this was 
entirely determined by the toron field contribution. In four dimensions, 
however, we find that it is zero for toron field configurations.
This leaves the interesting possibility that the imaginary
part of the effective action is generally zero in this model.

%\section{CONCLUSIONS} 

\section*{ACKNOWLEDGEMENT} 

This work is supported in part by INTAS grant INTAS-96-370. 
Discussions with M. G\"ockeler about the Ginsparg-Wilson relation are 
gratefully acknowledged.

\end{document}